\documentclass[11pt,a4paper]{article}
\pdfoutput=1
\usepackage{jheppub}
	\usepackage{amsmath, amssymb} 
	\usepackage{amsfonts}
	\usepackage{dsfont}
	\usepackage[ansinew]{inputenc}
	\usepackage[usenames,dvipsnames]{pstricks}
	\usepackage{epsfig}
	\usepackage{pst-grad} 
	\usepackage{pst-plot} 
	\usepackage{hyperref}
	\usepackage{graphicx}
	\usepackage{caption} 
	\usepackage{subcaption}
	\hypersetup
	{
		colorlinks,%
		citecolor=black,%
		linkcolor=black,%
		urlcolor=black,%
	}

\newcommand{\be}{\begin{equation}}
\newcommand{\ee}{\end{equation}}

	\newcommand{\vev}[1]{{\left< {#1} \right>}}
	
\title{On scalar radiation}

\author{Bartomeu Fiol and}
\author{Jairo Mart\'inez-Montoya}

\affiliation{Departament de Física Qu\`antica i Astrof\'isica, Institut de Ci\`encies del Cosmos (ICCUB), \\
Universitat de Barcelona, Mart\'i i Franqu\`es 1, 08028 Barcelona, Spain}

\emailAdd{bfiol@ub.edu}
\emailAdd{jmartinez@icc.ub.edu}

\abstract{
We discuss radiation in theories with scalar fields. Our key observation is that even in flat spacetime, the radiative fields depend qualitatively on the coupling of the scalar field to the Ricci scalar: for non-minimally coupled scalars, the radiative energy density is not positive definite, the radiated power is not Lorentz invariant and it depends on the derivative of the acceleration. We explore implications of this observation for radiation in conformal field theories.  First, we find a relation between two coefficients that characterize radiation, that holds in all the conformal field theories we consider. Furthermore, we find evidence that for a $1/2$-BPS probe coupled to ${\cal N}=4$ super Yang-Mills, and following an arbitrary trajectory, the spacetime dependence of the one-point function of the energy-momentum tensor is independent of the Yang-Mills  coupling.
}

\begin{document}
\maketitle


\section{Introduction}
The study of the creation and propagation of field disturbances by sources is one of the basic questions in any field theory. In classical electrodynamics, emission of electromagnetic waves by charged particles is of paramount importance, both at the conceptual and practical level \cite{rohrlich}. Similarly, the recent detection of gravitational waves \cite{Abbott:2016blz} provides a striking confirmation of General Relativity, and opens a new way to explore the Universe.

Understandably, radiation of massless scalar fields due to accelerated probes coupled to them, has received much less attention \cite{Cawley:1969re}. An exception is the study of radiation in scalar-tensor theories of gravity, since the radiation pattern can differ from General Relativity \cite{Will:2014kxa}.

The comments above refer to classical field theories. Recent formal developments, like holography and supersymmetric localization, have allowed to explore radiation in the strong coupling regime of conformal field theories (CFTs), which if they admit a Lagrangian formulation, very often include scalar fields. Some of the results of these explorations are, however, unexpected and even conflicting, as we now review.

In field theory, radiation is determined by the one-point function of the energy-momentum tensor of the field theory in the presence of an accelerated probe, which is described by a Wilson line $W$,
\be
\vev{T_{\mu \nu}}_W=\frac{\vev{W T_{\mu \nu}}}{\vev{W}}
\ee
Instead of computing $\vev{T_{\mu \nu}}_W$ for arbitrary trajectories, one can consider particularly simple kinematical configurations. A first possibility is motion with constant proper acceleration. The reason for this choice is that in any CFT,  a special conformal transformation maps a worldline with constant proper acceleration to a static one, for which $\vev{T_{\mu \nu}}_W$ is fixed up to a coefficient  \cite{Kapustin:2005py}
\be
\left.\frac{\vev{W T^{00}}}{\vev{W}} \right\vert_{\vec v=0}= \frac{h}{|\vec x|^4}
\label{defofh}
\ee
where $|\vec x|$ is the distance between the static Wilson line, placed at the origin, and the point where the measure takes place. The coefficient $h$ should thus capture the radiated power, at least for a probe with constant proper acceleration \cite{Fiol:2012sg}. 

A second interesting kinematical situation is that of the probe receiving a sudden kick. The Wilson line associated to the probe exhibits a cusp, and its vacuum expectation value develops a divergence, characterized by the cusp anomalous dimension \cite{Polyakov:1980ca} $\Gamma(\varphi)$, that depends on the rapidity of the probe after the kick. The expansion of $\Gamma(\varphi)$ for small $\varphi$,
\be
\Gamma(\varphi)= B\varphi^2+\dots
\label{defofb}
\ee
defines the Bremsstrahlung function $B$ \cite{Correa:2012at}. It was argued in \cite{Correa:2012at} that this function determines the energy radiated by a probe coupled to a CFT, since it appears in 
\be
E=2\pi B \int dt (\vec a)^2 
\label{raden}
\ee
If one grants this relation and further assumes that for arbitrary CFTs the radiated power is Lorentz invariant, one arrives at a Larmor-type formula
\be
{\cal P}=-2\pi B a^\lambda a_\lambda
\label{cftpower}
\ee
where $a^\lambda$ is the 4-acceleration. It was further argued in \cite{Correa:2012at} that in any CFT, the Bremsstrahlung function is universally related to the coefficient $C_D$ of the 2-point function of the displacement operator of any line defect \cite{Billo:2013jda}, by $12 B=C_D$. For Lagrangian CFTs with ${\cal N}=2$ supersymmetry this function can be computed using supersymmetric localization \cite{Fiol:2012sg, Fiol:2015spa, Fiol:2018yuc}. For ${\cal N}=2$ SCFTs it was argued \cite{Lewkowycz:2013laa, Fiol:2015spa} and then proved \cite{Bianchi:2018zpb} that $B=3h$. This relation is not satisfied in Maxwell's theory \cite{Lewkowycz:2013laa}, proving that no universal relation between $B$ and $h$ exists that is valid for all CFTs. 

Turning to holography, radiation by accelerated charges in a CFT is studied by first introducing a holographic probe, a string or a D-brane. Computations can be done at the worldsheet/worldvolume level, or taking into account the linear response of the gravity solution due to the presence of the holographic probe. Intriguingly, these two methods do not fully agree. At the holographic probe level, the computation of \cite{Mikhailov:2003er}, followed by \cite{Fiol:2011zg, Fiol:2014vqa} indicated that for a $1/2$-BPS probe coupled to ${\cal N}=4$ super Yang-Mills, in the large $N$, large $\lambda$ limit, the total radiated power is indeed of the form given by (\ref{cftpower}). The beautiful works \cite{Athanasiou:2010pv, Hatta:2011gh} dealt with the backreacted holographic computations, see also \cite{Hubeny:2010bq, Baier:2011dh, Agon:2014rda}.  The work \cite{Athanasiou:2010pv} considered only a probe in circular motion,  and found agreement with (\ref{cftpower}). However, the work \cite{ Hatta:2011gh} dealt with arbitrary trajectories, and found
\be
{\cal P}=-2\pi B\left(a^\lambda a_\lambda+\frac{1}{9}\frac{\dot a^0}{\gamma}\right)
\label{hattapower}
\ee
where $\gamma$ is the usual Lorentz factor. The additional term in (\ref{hattapower}) would imply that the radiated power in ${\cal N}=4$ SYM is not Lorentz invariant. The work \cite{Athanasiou:2010pv} was restricted to circular motion in a particular frame where $\dot a^0=0$, so by construction, it was not sensitive to the presence of the additional term in (\ref{hattapower}).

The angular distribution of radiated power is a more refined quantity than the total radiated power. At strong coupling it has been studied in \cite{Athanasiou:2010pv, Hatta:2011gh}, where the angular distribution of radiation emitted by a $1/2$-BPS probe coupled to ${\cal N}=4$ super Yang-Mills was determined holographically. Some of the features of the angular distribution of radiation found in \cite{Athanasiou:2010pv, Hatta:2011gh} were unexpected, like regions with negative energy density, or its dependence on the derivative of the acceleration, eq. (\ref{hattapower}). This prompted  \cite{Hatta:2011gh} to consider them artifacts of the supergravity approximation.

In this work we revisit the issue of radiation in scalar field theory, bringing new insights to many of the issues reviewed above. Our key observation is rather elementary: scalar fields couple to the scalar curvature of spacetime via the term \cite{Birrell:1982ix} $\xi R \phi^2$  so, even in flat spacetime, the energy-momentum tensor \cite{Callan:1970ze} and therefore the pattern of radiation, depend on $\xi$. In particular, radiation in conformal field theories requires considering conformally coupled scalars ($\xi=1/6$) instead of minimally coupled ones, $\xi=0$, as done in the field theory computations of \cite{Athanasiou:2010pv, Hatta:2011gh}.

Once we take this observation into account, we find that already at the level of free theory, radiation for a free conformal scalar displays the features that were found holographically for ${\cal N}=4$ super Yang-Mills: the radiated power is not Lorentz invariant, it depends on $\dot a$ and the radiated energy density is not everywhere positive. We conclude that these are generic features valid for all conformal field theories that include conformal scalars. In particular, eqs. (\ref{raden}) and (\ref{cftpower}) are not valid for arbitrary trajectories in CFTs with scalar fields.

Our observation also brings a new perspective to the lack of a universal relation between the coefficients $B$ and $h$ discussed above. In \cite{rohrlichnc,rohrlich} a manifestly Lorentz invariant quantity, the invariant radiation rate ${\cal R}$, was defined in the context of Maxwell theory. We extend the definition, and show that while in Maxwell theory ${\cal R}={\cal P}$, this is not true in general CFTs. For the probes and CFTs considered in this work, ${\cal R}$ can be written as
\be 
{\cal R}=-2\pi B_{\cal R} a^\lambda a_\lambda
\ee
where $B_{\cal R}$ is a new coefficient that in general differs from the Bremsstrahlung function $B$. Furthermore we find that the relation
\be 
B_{\cal R}=\frac{8}{3}h
\ee
holds in all the cases considered. This relation has thus the potential to be universal for all probes and all CFTs.

We turn then our attention to Lagrangian ${\cal N}=2$ SCFTs, and for ${\cal N}=4$ super Yang-Mills, we do find a surprise. The full one-point function of the energy density in the presence of a probe following an arbitrary trajectory has exactly the same spacetime dependence at weak and at strong 't Hooft coupling. This leads us to conjecture that this quantity is protected by non-renormalization. This would be rather surprising, as for generic timelike trajectories, $\vev{T^{\mu \nu}}_W$ is not a BPS quantity.

The structure of the paper is the following. In section \ref{freetheories}, we revisit radiation by probes coupled to free field theories. We show that once we take into account the improvement term of the energy-momentum tensor for non-minimally coupled scalars, the radiative energy density is not positive definite, which is just a manifestation of the more general fact that non-minimally coupled scalars can violate energy conditions even classically \cite{Bekenstein:1974sf}. Furthermore, for non-minimally coupled scalars, the radiated power ${\cal P}$ is not Lorentz invariant.  The new term that we find in the rate of 4-momentum loss is formally similar to the Schott term  that appears in the Lorentz-Dirac equation in electrodynamics \cite{rohrlich}. We will argue however that in theories with non-minimally coupled scalars its origin and meaning are different than the Schott term in classical electrodynamics.

In section \ref{onepointcft}, we discuss constraints imposed by conformal symmetry on the one-point function of the energy-momentum tensor of a conformal field theory, in the presence of an arbitrary timelike line defect.

In section \ref{susyradiation} we discuss radiation by $1/2$-BPS probes coupled to ${\cal N}=2$ SCFTs. Quite remarkably, for a $1/2$-BPS probe coupled to ${\cal N}=4$ super Yang Mills following an arbitrary trajectory, the classical computation with conformally coupled scalars matches exactly the angular distribution found holographically  \cite{Athanasiou:2010pv, Hatta:2011gh}. 

In section \ref{outlook} we mention some open questions. Our conventions are as follows: we work with a mostly minus metric, so the 4-velocity $u$ and the 4-acceleration $a$ satisfy $u^2=1, a^2<0$. Dots have different meaning for vectors and 4-vectors: $\dot a=da/d\tau$, but $\dot{\vec a}= d \vec a/dt$. Our overall normalization of the energy-momentum tensor for scalars is not the usual one; it has been chosen for convenience when we add scalar and vector contributions in supersymmetric theories.

\section{Radiation in free field theories}
\label{freetheories}
Consider a probe coupled to a field theory, following an arbitrary, prescribed, timelike trajectory $z^\mu(\tau)$. One first solves the equations of motion for the field theory, in the presence of this source, choosing the retarded solution. Let $x^\mu$ be the point where the field is being measured; define $\tau_{ret}$ by the intersection of the past light-cone of $x^\mu$ and the worldline of the probe, and the null vector $\ell=x-z(\tau_{ret})$. 

One then evaluates the energy-momentum tensor with the retarded solution. Usually one defines the radiative part of the energy-momentum tensor $T^{\mu \nu}_r$ as the piece that decays as $1/r^2$  so it yields a nonzero flux arbitrarily far away from the source. A more restrictive definition of $T^{\mu \nu}_r$ was introduced in \cite{Teitelboim:1970mw, Teitelboim:1979px}, who required that 
\begin{itemize}
\item{$\partial _\mu T^{\mu \nu}_r=0$ away from the source.}
\item{$\ell_\mu T^{\mu \nu}_r=0$ so flux through the light-cone emanating from the source is zero.}
\item{$T^{\mu \nu}_r =\frac{A}{(\ell \cdot u)^4}\ell^\mu \ell^\nu$ with $A$ a Lorentz scalar.}
\item{$A\geq 0$ so the radiative energy density is nonnegative.}
\end{itemize}
In this work we will consider theories that don't satisfy the weak energy condition classically; for these theories, the requirement that the radiative energy density is nonnegative is less well motivated. In this work we use the first definition of $T^{\mu \nu}_r$, but we will discuss the implications of considering the second one. From $T^{\mu \nu}_r$ we define \cite{rohrlich}
\be 
\frac{dP^\mu}{d\tau d \Omega}= r^2 T^{\mu \nu}_r u_\nu
\ee
and integrating over the solid angle we obtain $dP^\mu/d\tau$. It is a 4-vector \cite{schild} that gives the rate of energy and momentum emitted by the probe. From it one can define two quantities. The first one is the radiated power ${\cal P}$,
\be
{\cal P}=\frac{dP^0}{dt}
\ee
which is not manifestly Lorentz invariant. Following Rohrlich \cite{rohrlich}, we define a second quantity, the invariant radiation rate ${\cal R}$ as
\be 
{\cal R}=u_\mu \frac{dP^\mu}{d\tau}
\ee
which is manifestly Lorentz invariant. For free CFTs, this invariant radiation rate can be written as
\be
{\cal R}=-2\pi B_{\cal R} a^\lambda a_\lambda
\label{brfree}
\ee
We don't have a proof that this is the most generic form that ${\cal R}$ can take in interacting CFTs, but let's mention some restrictions. In principle there could be also a term in (\ref{brfree}) proportional $u\cdot \dot a$, but since $a^2=-u\cdot \dot a$, it would be redundant. Furthermore, by dimensional analysis, terms with higher derivatives of $a$ can't appear in (\ref{brfree}). In conclusion, (\ref{brfree}) is the most general form that ${\cal R}$ can take, if it depends only on Lorentz invariants evaluated at a single retarded time.

\subsection{Maxwell field}
The energy-momentum tensor is
\be
T^{\mu \nu}=\frac{1}{4\pi}\left(F^{\mu \lambda}F_\lambda^{\, \, \nu}+\frac{1}{4}\eta^{\mu \nu}F_{\alpha \beta}F^{\beta \alpha}\right)
\ee
It is traceless, without using the equations of motion. Consider a probe coupled to the Maxwell field, with charge $q$, following an arbitrary trajectory. The full energy-momentum tensor evaluated on the retarded solution is \cite{schild}
\be 
T^{\mu \nu}=\frac{q^2}{4\pi}\left(\frac{\ell^\mu u^\nu+\ell^\nu u^\mu}{(\ell\cdot u)^5}(1-\ell\cdot a)+\frac{\ell^\mu a^\nu+\ell^\nu a^\mu}{(\ell\cdot u)^4}-\frac{a^2}{(\ell\cdot u)^4}\ell^\mu \ell^\nu-\frac{(1-\ell \cdot a)^2}{(\ell\cdot u)^6}\ell^\mu \ell^\nu-\frac{1}{2}\frac{\eta^{\mu \nu}}{(\ell\cdot u)^4}\right)
\label{vectortmunu}
\ee
where all quantities are evaluated at retarded time. Evaluating (\ref{vectortmunu}) for a static probe we derive the $h$ coefficient \cite{Kapustin:2005py}
\be 
\left. T^{00}\right\vert_{\vec v=\vec 0}= \frac{q^2}{8\pi}\frac{1}{r^4} \hspace{.5cm}\Rightarrow \hspace{.5cm} h=\frac{q^2}{8\pi}
\label{hforem}
\ee
The part of (\ref{vectortmunu}) decaying as $1/r^2$ is
\be
T^{\mu \nu}_r=-\frac{q^2}{4\pi}
\left(\frac{a^2}{(\ell \cdot u)^4}+\frac{(\ell \cdot a)^2}{(\ell \cdot u)^6}\right) \ell^\mu \ell^\nu
\label{radvectortmunu}
\ee
It satisfies all the criteria of  \cite{Teitelboim:1970mw, Teitelboim:1979px}, so it is the radiative part according to both definitions. Integration over angular variables yields
\be 
\frac{dP^\mu}{d\tau}=-\frac{2}{3}q^2 a^\lambda a_\lambda u^\mu 
\ee
It is a future-oriented timelike 4-vector, guaranteeing that all inertial observers agree that the particle is radiating away energy. The relativistic Larmor's formula follows 
\be
{\cal P}={\cal R}=-\frac{2}{3}q^2 a^\lambda a_\lambda
\label{larmor}
\ee
recall that $a^2<0$ in our conventions. From (\ref{larmor}) we derive the Bremsstrahlung coefficient for Maxwell's theory,
\be
B=\frac{q^2}{3\pi}
\label{bofem}
\ee
It follows from (\ref{hforem}) and (\ref{bofem}) that  \cite{Lewkowycz:2013laa}
\be
B=\frac{8}{3}h
\label{bhratioinem} .
\ee

\subsection{Scalar fields}
Consider a free massless scalar field, with arbitrary coupling $\xi$ to the Ricci scalar. The energy-momentum tensor is \cite{Callan:1970ze}
\be
4\pi T^{\mu \nu}=\partial^\mu \phi \partial^\nu \phi-\frac{1}{2}\eta^{\mu \nu} \partial_\alpha \phi \partial^\alpha \phi-\xi (\partial^\mu \partial^\nu-\eta^{\mu \nu}\square)\phi^2
\label{improvedtmunu}
\ee
In general, the trace of (\ref{improvedtmunu}) does not vanish, even when applying the equations of motion. For the conformal value $\xi=\frac{1}{6}$ it vanishes away from the sources, if we apply the equations of motion. For $\xi \neq 0$, this energy-momentum tensor  can violate the weak energy condition at the classical level \cite{Bekenstein:1974sf}, even in Minkowski space.

Now consider a probe coupled to the scalar field, following an arbitrary trajectory. The energy-momentum tensor (\ref{improvedtmunu}) evaluated on the retarded solution of the equation of motion is 
\begin{multline}
4\pi T^{\mu \nu}= \frac{q^2}{(\ell\cdot u)^4} \left((1-6\xi)u^\mu u^\nu-(1-8\xi)\frac{1-\ell \cdot a}{\ell\cdot u}(\ell^\mu u^\nu + \ell^\nu u^\mu)+2\xi (\ell^\mu a^\nu+\ell^\nu a^\mu)\right . \\
+\left. (1-8\xi) \frac{(1-\ell\cdot a)^2}{(\ell\cdot u)^2}\ell^\mu \ell^\nu +2\xi \frac{\ell \cdot \dot a}{\ell \cdot u}\ell^\mu \ell^\nu+\frac{1-8\xi}{2}\eta^{\mu \nu} -(1-6\xi)(\ell \cdot a)\eta^{\mu \nu} \right)
\label{scalartmunu}
\end{multline}
evaluated at retarded time. It depends on $\dot a= da/d\tau$, because the improved energy-momentum tensor (\ref{improvedtmunu}) involves second derivatives of the field, and the solution depends on the velocity of the probe. 

In the conformal case $\xi=1/6$ the terms independent or linear in the acceleration are the same as in (\ref{vectortmunu}), up to an overall factor. In the next section, we will argue that these terms are actually universal for all CFTs.

Evaluating (\ref{scalartmunu}) on a static probe for the conformal value $\xi=1/6$, we derive \cite{Kapustin:2005py}
\be 
\left. T^{00}\right\vert_{\vec v=\vec 0}=\frac{1-4\frac{1}{6}}{8\pi}\frac{q^2}{r^4} \, \, \Rightarrow \, \,  h=\frac{1}{24 \pi}q^2
\ee


The part of (\ref{scalartmunu}) decaying as $1/r^2$ is
\be 
T^{\mu \nu}_r = \frac{q^2}{4\pi} \left( (1-8\xi)\frac{(\ell\cdot a)^2}{(\ell\cdot u)^6}+2\xi \frac{\ell\cdot \dot a}{(\ell\cdot u)^5}\right) \ell^\mu \ell^\nu
\label{radscalartmunu}
\ee

It satisfies the first three criteria of \cite{Teitelboim:1979px} to be the radiative part. It also satisfies $|T^{00}|=|T^{0i}|$.  As a check, for $\xi=0$, it reduces to the energy density found in \cite{Athanasiou:2010pv}, which is manifestly positive definite. However, for $\xi\neq 0$, $T^{00}$ is not guaranteed to be positive. After integration over the angular variables, we find
\be
\frac{dP^\mu}{d\tau}=-\frac{1}{3}q^2 a^\lambda a_\lambda u^\mu-\frac{2\xi}{3}q^2 \dot a^\mu
\label{scalarrate}
\ee
The improvement term in the energy-momentum tensor of the scalar field (\ref{improvedtmunu}) induces a qualitatively new term in $dP^\mu/d\tau$, compared with the electrodynamics case. The
additional term in (\ref{scalarrate}) is a total derivative, and it is formally identical to the Schott term in classical electrodynamics \cite{rohrlich}. However, the origin is different. In classical electrodynamics, the Schott term appears in the Lorentz-Dirac equation of motion of the probe, and it can be deduced from the fields created by the probe, in the zone near its worldline. It does not appear from evaluating the radiative part of the energy-momentum tensor (\ref{radvectortmunu}). On the other hand,  in (\ref{scalarrate}) the new term appears directly from evaluating the energy-momentum tensor of the fields that decay like $1/r^2$, away from the probe.

This additional term that we have encountered  in (\ref{scalarrate}) in a free theory computation has the same form as the additional term found holographically by \cite{Hatta:2011gh}, eq. (\ref{hattapower}). In that context, the works \cite{Baier:2011dh, Agon:2014rda} have advocated using the more restrictive definition of $T^{\mu \nu}_r$, thus setting $\xi=0$ in (\ref{radscalartmunu}, \ref{scalarrate}). An argument in favor of doing so 
is that the new term in (\ref{scalarrate}) is a total derivative so, for instance, its contribution vanishes for any periodic motion when integrated over a full period. This clashes with the intuition of radiated energy as something irretriavably lost by the particle.  However, we think this intuition is built on the idea that the energy density is positive definite, which is not the case for non-minimally coupled fields.

For a minimally coupled scalar field, $\xi=0$, ${dP^\mu}/{d\tau}$ is again a future-oriented, timelike 4-vector, and ${\cal P}={\cal R}$, as in Maxwell's theory \cite{Cawley:1969re, Athanasiou:2010pv}. On the other hand, for $\xi \neq 0$, this 4-vector is no longer guaranteed to be timelike. This is related with $T^{00}$ no longer being positive definite. In the instantaneous rest frame,
\be 
\left.\frac{dP^\mu}{d\tau}\right\vert_{\vec v=0}=\left( \frac{1-2\xi}{3}q^2 \vec a^2,-\frac{2\xi}{3}q^2 \dot {\vec a}\right)
\ee
So for $\xi<1/2$, in the instantaneous rest frame, there is energy loss. However, if $dP^\mu/d\tau$ is spacelike, the sign of its zeroth component is no longer the same in all inertial frames.

For a non-minimally coupled scalar, ${\cal P}$ and ${\cal R}$ no longer coincide, and ${\cal P}$ is not Lorentz invariant. Indeed,
\be
{\cal P}= -\frac{1}{3}q^2 a^\lambda a_\lambda-\frac{2\xi}{3} q^2\frac{\dot a^0}{\gamma}
\label{xipower}
\ee
and
\be
{\cal R}=-\frac{1-2\xi}{3}q^2 a^\lambda a_\lambda
\ee
For non-minimally coupled scalars, we will still define $2\pi B$ as the coefficient in front of the $-a^\lambda a_\lambda$ term in (\ref{xipower}). We furthermore introduce a new coefficient $B_\xi$, as the coefficient in ${\cal R}=-2\pi B_\xi a^\lambda a_\lambda$. We obtain
\be
B_\xi=\frac{1-2\xi}{6\pi}q^2
\ee
Notice that $B_{\xi=0}=B$; we also define $B_{{\cal R}}=B_{\xi=1/6}$. In particular, for the conformally coupled scalar it follows that $B_{{\cal R}}=\frac{8}{3} h$. This ratio is the same as in Maxwell's theory, eq. (\ref{bhratioinem}). 

\section{One-point function of the energy-momentum tensor in CFTs.}
\label{onepointcft}
In this section we discuss the constraints that conformal invariance imposes on the one-point function of the energy-momentum tensor of a conformal field theory, in the presence of a timelike line defect. While in the rest of the paper we consider Lagrangian field theories and the line defects are Wilson lines, the arguments of this section apply to arbitrary line defects in general CFTs.

For classical conformal field theories, we have seen in the previous section that the full one-point function of the energy-momentum tensor at a point in spacetime depends on the value of the 4-velocity and the 4-acceleration evaluated at a single retarded time. It is far from obvious that this feature should hold for generic line defects in arbitrary CFTs. In fact, once one considers strongly coupled conformal non-Abelian gauge theories, there are compelling arguments \cite{Hatta:2011gh} that virtual timelike quanta will decay into further quanta thus forming a cascade, so the radiation measured at a point in spacetime does not have its origin at just a single retarded time in the probe worldline. This picture suggests that at least in some theories, the full one-point function should include integrals over the worldline of the probe, up to the retarded time,
\be
\vev{T^{\mu \nu}}_W = \int^{\tau_{ret}} d\tau f(a) +\dots
\label{memory}
\ee
to take into account radiation originated by the cascade of timelike virtual quanta. Intriguingly enough, the holographic computations of \cite{Athanasiou:2010pv, Hatta:2011gh} do not find such terms for ${\cal N}=4$ SYM in the planar limit. We will make a small comment about the presence or not of these terms for generic CFTs at the end of this section. 

In the present discussion we will focus on the terms where the kinematic 4-vectors, like the 4-velocity and the 4-acceleration appear in the answer evaluated at a single time, without any integrals. Dimensional analysis, conformal symmetry and conservation of the energy-momentum tensor constraint the form of the answer.

The full energy-momentum tensor of a CFT in the presence of a static probe is fixed by conformal invariance \cite{Kapustin:2005py}, up to an overall coefficient,
\be
\vev{T^{00}}_W = \frac{h}{|\vec x|^4} \hspace{.5cm} \vev{T^{0i}}_W = 0 \hspace{.5cm} \vev{T^{ij}}_W = \frac{h}{|\vec x|^4} \left(\delta^{ij}-2 \frac{x^i x^j}{|\vec x|^2} \right)
\ee
By applying a boost, it is then also fixed for a probe with constant velocity. This determines all the acceleration independent terms; since they are universal, they can be read off from (\ref{vectortmunu}) or (\ref{scalartmunu}). These terms decay as $1/r^4$ as dictated by dimensional analysis,
\be
\left.\vev{T^{\mu \nu}}_W\right\vert_{\vec v=constant} = h \left( -\frac{\eta^{\mu \nu}}{(\ell \cdot u)^4}+2 \frac{\ell^\mu u^\nu+\ell^\nu u^\mu}{(\ell \cdot u)^5}-2\frac{\ell^\mu \ell^\nu}{(\ell \cdot u)^6} \right)
\ee

Furthermore, by applying a special conformal transformation to a static worldline, one obtains a worldline with constant proper acceleration. Therefore, for any CFT, the full energy-momentum tensor for a hyperbolic line defect is completely determined up to an overall constant. It is immediate to check that $\vev{T_{\mu \nu}}_W$ for Maxwell theory, eq. (\ref{vectortmunu}), and  for a conformal scalar, eq. (\ref{scalartmunu}) with $\xi=1/6$, have the same spacetime dependence for hyperbolic motion, since in this case $\dot a =-a^2 u$.

We will now argue that the previous property implies that the terms linear in the 4-acceleration $a$ must also be universal. The argument goes as follows. Since a worldline with constant proper acceleration satisfies $\dot a =-a^2 u$, terms that are not universal in $T^{\mu \nu}$ and change from one CFT to another, must be such that they collapse to the same universal expression when $\dot a =-a^2 u$. But terms linear in $a$ don't depend on $\dot a$ or $a^2$, so they must be universal for all CFTs. These terms decay as $1/r^3$ as dictated by dimensional analysis. All in all, the terms independent or linear in $a$ are,
\be
\left.\vev{T^{\mu \nu}}_W \right\vert_{{\cal O}(a)}= 2 h \left( -\frac{1}{2} \frac{\eta^{\mu \nu}}{(\ell \cdot u)^4}+(1-\ell\cdot a) \frac{\ell^\mu u^\nu+\ell^\nu u^\mu}{(\ell \cdot u)^5}+\frac{\ell^\mu a^\nu+\ell^\nu a^\mu}{(\ell \cdot u)^4}-(1-2\ell\cdot a) \frac{\ell^\mu \ell^\nu}{(\ell \cdot u)^6} \right)
\label{universalpiece}
\ee
We then conclude that the terms in $\vev{T_{\mu \nu}}_W$ independent or linear in the 4-acceleration $a^\lambda$ - which respectively decay as $1/r^4$ and $1/r^3$ - are universal for all CFTs. On the other hand, terms that involve $a^2$ or $\dot a$ and decay like $1/r^2$ are not uniquely fixed by conformal invariance. Indeed, the $1/r^2$ terms  for Maxwell's theory (\ref{radvectortmunu}) and a conformal scalar (\ref{radscalartmunu}) are different.

The formula (\ref{universalpiece}) refers only to terms that depend only on the probe worldline at the retarded time, and does not exclude potential additional terms of the schematic form (\ref{memory}). To conclude this section, let's comment on the restrictions that conservation of the energy-momentum tensor imposes on the presence of possible terms of the type (\ref{memory}), that depend on the worldline of the probe, and not just the retarded time. First of all, the full energy-momentum tensor is conserved. We can further require that the piece of the energy-momentum tensor that decays like $1/r^2$ is conserved by itself, since it corresponds to energy that is detached from the probe. It then follows that the piece of $\vev{T^{\mu \nu}}$ that doesn't decay like $1/r^2$ must also be conserved by itself. It is straightforward to check that the terms that appear explicitly in (\ref{universalpiece}) are conserved. This implies that if there are additional terms of the type (\ref{memory}) that decay like faster than $1/r^2$ beyond the ones that appear in (\ref{universalpiece}), they must be conserved on their own.

\section{Radiation in ${\cal N}=2$ superconformal theories}
\label{susyradiation}
The discussion in the previous section was completely classical. In this section we consider ${\cal N}=2$ Lagrangian SCFTs, for which powerful techniques to study the strong coupling regime are available.

Consider the energy-momentum tensor created by a $1/2$-BPS probe coupled to a Lagrangian ${\cal N}=2$ SCFT in the classical limit. The probe is coupled to a vector and a scalar in the adjoint representation of the gauge group. As argued in  \cite{Athanasiou:2010pv, Hatta:2011gh}, at very weak coupling this amounts to adding the contribution of the Maxwell (\ref{radvectortmunu}) and free scalar (\ref{radscalartmunu}) terms, with an effective charge. However \cite{Athanasiou:2010pv, Hatta:2011gh} considered a free minimally coupled scalar. In CFTs, the correct computation amounts to adding (\ref{vectortmunu}) and (\ref{scalartmunu}) with the conformal value, $\xi=1/6$. We obtain

\begin{multline} 
T^{\mu \nu}_{{\cal N}=2}=
2 h^{\mathcal{N}=2} \left( -\frac{1}{2} \frac{\eta^{\mu \nu}}{(\ell \cdot u)^4}+(1-\ell\cdot a) \frac{\ell^\mu u^\nu+\ell^\nu u^\mu}{(\ell \cdot u)^5}+\frac{\ell^\mu a^\nu+\ell^\nu a^\mu}{(\ell \cdot u)^4}-(1-2\ell\cdot a) \frac{\ell^\mu \ell^\nu}{(\ell \cdot u)^6}\right) \\
+\frac{h^{\mathcal{N}=2}}{2}
\left( -\frac{3a^2}{(\ell \cdot u)^4}+\frac{\ell\cdot \dot a}{(\ell \cdot u)^5}-4\frac{(\ell\cdot a)^2}{(\ell \cdot u)^6}\right)\ell^\mu \ell^\nu
\label{tmunususy}  
\end{multline}

In three-dimensional language, with $\vec n= \frac{\vec r-\vec z}{|\vec r-\vec z|}$, the radiative energy density is
\be 
T^{00}_{{\cal N}=2}=
\frac{h^{\mathcal{N}=2}}{2 r^2 }\left(
\frac{4\left|\vec a\right|^2+3\gamma^2 (\vec \beta\cdot \vec a)^2+\vec \beta\cdot \dot{\vec a}}{(1-\vec \beta \cdot \vec n)^4}
+\frac{5  (\vec \beta \cdot \vec a)(\vec n \cdot \vec a) -\gamma^{-2} \vec n\cdot \dot{\vec a}}{(1-\vec \beta \cdot \vec n)^5}
-4\frac{\gamma^{-2}(\vec n\cdot \vec a)^2}{(1-\vec \beta \cdot \vec n)^6}
\right) 
\label{t00susy}
\ee
Our free classical computation only guarantees (\ref{tmunususy}, \ref{t00susy}) at leading order in $\lambda$, for small $\lambda$. Strikingly, the 00 component of (\ref{tmunususy}) is exactly the same result found by a rather elaborate holographic computation for a $1/2$-BPS probe in the fundamental representation of ${\cal N}=4$ SU($N$) super Yang-Mills in \cite{Athanasiou:2010pv, Hatta:2011gh}, in the planar limit and at strong 't Hooft coupling where \cite{Mikhailov:2003er} $3h=B=\sqrt{\lambda}/4\pi^2$ ! To elaborate, we have computed the $1/r^4, 1/r^3$ terms at strong coupling, using the results of the holographic computations of \cite{Athanasiou:2010pv, Hatta:2011gh} and have found exactly the first line of (\ref{tmunususy}). The match of the spacetime dependence of these terms at weak and strong coupling is not surprising, as we have argued in section \ref{onepointcft} that they are universal. Nevertheless, this match does provide a strong check of the holographic computations in \cite{Athanasiou:2010pv, Hatta:2011gh}. On the other hand, the $1/r^2$ term (\ref{t00susy}) was already computed at strong coupling in \cite{Athanasiou:2010pv, Hatta:2011gh}, and again it displays the same spacetime dependence as the classical result.  We stress that we find exact agreement at the level of energy density, before performing any time average. This agreement prompts us to conjecture that (\ref{tmunususy}) is true for all values of $\lambda$, in the planar limit. It is tempting to conjecture that (\ref{tmunususy}) is true even at finite $N$ and finite $\lambda$, but we currently don't have evidence for this stronger claim. Conformal symmetry alone is not enough to explain this agreement: comparing (\ref{radvectortmunu}), (\ref{radscalartmunu}) and (\ref{tmunususy}) it is clear that the radiative energy density of a probe in arbitrary motion is not the same for different conformal field theories. Furthemore, while the probe is $1/2$-BPS, it is following an arbitrary trajectory, so the Wilson line does not preserve any supersymmetry globally.

Many of the unexpected features of (\ref{t00susy}) have simple classical explanations that arise from properties of conformally coupled scalars: the fact that (\ref{t00susy}) is not positive definite everywhere, was interpreted in \cite{Athanasiou:2010pv} as an inherently quantum effect. In fact, it's a feature already present at the classical level, reflecting that conformally coupled scalar fields can violate energy conditions even classically. As first noticed in  \cite{Hatta:2011gh},  (\ref{t00susy}) depends on the derivative of the acceleration; now we understand that this follows from the fact that the improved tensor (\ref{improvedtmunu}) involves second derivatives of the field. Another puzzle raised in  \cite{Hatta:2011gh} is that in ${\cal N}=4$ SYM, radiation was isotropic at weak coupling; as our classical derivation of  (\ref{t00susy}) shows, this isotropy is just an artifact of considering minimally coupled scalars, instead of conformally coupled ones.

In \cite{Athanasiou:2010pv} it was noticed that for circular motion, while the angular distribution of radiated power computed holographically did not match the classical computation of Maxwell plus minimally coupled scalar, the respective time averages over a period did match. The reason is now easy to understand:  the details of the angular distribution depend on $\xi$, but after averaging over a period, the averaged angular distribution is independent of $\xi$. 

Let's discuss now the total radiated power in ${\cal N}=2$ SCFTs. Integration of (\ref{t00susy}) over angular variables yields
\be
\frac{dP^\mu}{d\tau}=-2\pi B^{\mathcal{N}=2} (a^\lambda a_\lambda u^\mu+\frac{1}{9} \dot a^\mu )
\label{susyrate}
\ee
Our computation ensures that this formula is valid at the classical level. At strong coupling, the only evidence is the ${\cal N}=4$ SYM holographic computation of \cite{Hatta:2011gh}. 

To conclude, let's comment on the relation $B^{{\cal N}=2}=3h^{{\cal N}=2}$ conjectured in \cite{Lewkowycz:2013laa, Fiol:2015spa} and proved in \cite{Bianchi:2018zpb} for generic, not necessarily Lagrangian, ${\cal N}=2$ SCFTs. This is a relation between the Bremsstrahlung coefficient as defined in (\ref{defofb}) and the $h^{\mathcal{N}=2}$ coefficient, as defined in (\ref{defofh}). The proof presented in \cite{Bianchi:2018zpb} relies on $12B^{\mathcal{N}=2}=C_D$, but not on the argument \cite{Correa:2012at} that identifies $B^{\mathcal{N}=2}$ defined in (\ref{defofb}) with the $h^{\mathcal{N}=2}$ coefficient in (\ref{defofh}). The values obtained in section \ref{freetheories} allow to test that this relation is satisfied by a free U(1) ${\cal N}=2$ SCFT, and in fact by any Lagrangian ${\cal N}=2$ SCFT at weak coupling,
\be 
B^{{\cal N}=2}=B^{EM}+ B^{scalar}= 3 (h^{EM}+h^{scalar})=3 h^{{\cal N}=2}
\ee
On the other hand, it also follows that the coefficients $B_{\cal R}^{\mathcal{N}=2}$ and $h^{\mathcal{N}=2}$ of any Lagrangian ${\cal N}=2$ SCFT satisfy, at weak coupling, the same relation as in Maxwell theory or for a conformal scalar,
\be 
B^{{\cal N}=2}_{\cal R}=B^{EM}_{\cal R}+ B_{\cal R}^{scalar}=\frac{8}{3}h^{EM}+\frac{8}{3}h^{scalar}=\frac{8}{3} h^{{\cal N}=2}
\ee
At strong coupling, contracting (\ref{susyrate}) with $u_\mu$ and using $B^{\mathcal{N}=2}=3h^{\mathcal{N}=2}$, we again obtain
\be
B_{{\cal R}}^{\mathcal{N}=2}=\frac{8}{9}B^{\mathcal{N}=2}=\frac{8}{3}h^{\mathcal{N}=2} 
\label{bccandh}
\ee
which is again the relation found for Maxwell's theory and for a free conformal scalar. So if (\ref{susyrate}) holds, (\ref{bccandh}) would be true for all the probes coupled to CFTs considered in this paper. Currently, the only evidence for (\ref{susyrate}) at strong coupling is the holographic computation of \cite{Hatta:2011gh} for ${\cal N}=4$ SYM.

\section{Discussion and outlook}
\label{outlook}
In this work we have discussed radiation for theories with scalar fields. We have found that for non-minimally coupled scalars, the energy density is no longer positive definite, it depends on the derivative of the acceleration of the probe, and the radiated power is not Lorentz invariant. These three features were also encountered in the strongly coupled regime of ${\cal N}=4$ super Yang-Mills, by holographic computations \cite{Athanasiou:2010pv, Hatta:2011gh}. In the introduction we mentioned that these computations do not quite agree with the holographic computations at the probe string/brane level. The backreacted computations of \cite{Athanasiou:2010pv, Hatta:2011gh} are on a firmer theoretical ground, but the results they yielded were unexpected, casting doubts on their validity. Our work implies that these features are to be expected for any conformal field theory with conformal scalars, and confirm the validity of the holographic computations of \cite{Athanasiou:2010pv, Hatta:2011gh}.


In this work we have not discussed radiation reaction on the probe coupled to the scalar field. It would be interesting to discuss it for the case of non-minimally coupled scalars.

We have shown that the relation (\ref{bccandh}) holds for probes of free CFTs, and we have presented evidence that it also holds for $1/2$-BPS probes in ${\cal N}=4$ SCFTs. At this point it is not clear whether it holds for arbitrary probes of generic CFTs. A possible case to further test it would be less supersymmetric probes of ${\cal N}=4$ super Yang-Mills. 

The fact that (\ref{t00susy}) holds both at weak and strong $\lambda$ in the planar limit of ${\cal N}=4$ super Yang-Mills is rather mysterious, as it is not a BPS quantity. It will be important to prove if  (\ref{t00susy}) holds for any $\lambda$, in the planar limit, or even at finite $N$. An even stronger conjecture is that it holds for generic ${\cal N}=2$ superconformal theories, but currently we lack techniques to study $\vev{T^{\mu \nu}}_W$ at strong coupling for generic ${\cal N}=2$ SCFTs and arbitrary timelike worldlines.

Finally, this note has only considered radiation of scalar fields in Minkowski spacetime. It will be interesting to generalize our results to other spacetimes.

\acknowledgments
We would like to thank Lorenzo Bianchi, Alberto G\"uijosa, Diego M. Hofman, Zohar Komargodski, Madalena Lemos, Hong Liu, Marco Meineri and Juan Pedraza for correspondence and comments on the draft. Research funded by Spanish MINECO under projects MDM-2014-0369 of ICCUB (Unidad de Excelencia "Mar\'ia de Maeztu") and FPA2017-76005-C2-P, and by AGAUR, grant 2017-SGR 754. J. M. M. is further supported by "la Caixa" Foundation (ID 100010434) with fellowship code LCF/BQ/IN17/11620067, and from the European Union's Horizon 2020 research and innovation programme under the Marie Sk{\l}odowska-Curie grant agreement No. 713673.

\end{document}